\documentclass[12pt]{article}
\usepackage[textures]{graphics}
\parskip=\bigskipamount
\newcommand{\beq}{\begin{equation}}
\newcommand{\eeq}{\end{equation}}
\newcommand{\beqa}{\begin{eqnarray}}
\newcommand{\eeqa}{\end{eqnarray}}
\newcommand{\ket} [1] {\vert #1 \rangle}
\newcommand{\bra} [1] {\langle #1 \vert}

\title{ Tomography of one and two qubit states and factorisation of the 
Wigner distribution in prime power dimensions.}
\date{}
\author{Thomas Durt \\{\footnotesize\it TENA TONA VUB 
Pleinlaan 2 1050 Brussels Belgium
\& thomdurt@vub.ac.be} }
\begin{document}

\maketitle
{\it We study different techniques that allow us to gain complete knowledge about an unknown quantum state, e.g. to perform full tomography 
of this state. We focus on two apparently simple cases, full tomography of
 one and two qubit systems. We analyze and compare those techniques according to two figures of merit. Our first criterion is the minimisation 
 of the redundancy of the data acquired during the tomographic process.
  In the case of two-qubits tomography, we also analyze this process from the point
   of view of factorisability, so to say we analyze 
 the possibility to realise
  the tomographic process through local operations and classical communications between
   local observers. This brings us naturally to study the possibility to factorize the (discrete)
    Wigner distribution of a composite system into the product of local Wigner distributions. The discrete Heisenberg-Weyl group is an essential ingredient of our approach. 
 Possible extensions of our results to higher dimensions are discussed
    in the last section and in the conclusions. }

\section{Introduction}

The estimation of an unknown state is one of the important problems in
quantum information and quantum computation\cite{Na1,N1}. Traditionnally,
 the estimation of the $d^2-1$ parameters that characterize the density matrix of a single 
qu$d$it consists of realising $d+1$ independent von Neumann measurements on the system.
 For instance, when the system is a spin 1/2 particle, three successive Stern-Gerlach measurements performed along orthogonal directions 
 make
  it possible to infer the values of the 3 Bloch parameters $p_{x},p_{y},$%
and $p_{z}$ defined by
\begin{equation}
\left\{ 
\begin{array}{l}
\left\langle \sigma _{x}\right\rangle =p_{x}=\gamma \sin \theta \cos
\varphi \\ 
\left\langle \sigma _{y}\right\rangle =p_{y}=\gamma \sin \theta \sin
\varphi \\ 
\left\langle \sigma _{z}\right\rangle =p_{z}=\gamma \cos \theta%
\end{array}%
\right.
\end{equation}
 Once we know the value of these parameters, we are able to determine unambiguously the 
 value of the density matrix, making use of the identity
  \begin{equation}
\rho (\gamma ,\theta ,\varphi )=\frac{1}{2}(I+p_{x}\sigma
_{x}+p_{y}\sigma _{y}+p_{z}\sigma _{z})=\frac{1}{2}(I+%
\overrightarrow{\gamma }.\overrightarrow{\sigma })
\end{equation}
When the qubit system is not a spin 1/2 particle but  consists of the polarisation of a photon, a similar result can
 be achieved by measuring its degree of polarisation in three independent polarisation bases, for instance with
  polarising beamsplitters, which leads to the Stokes representation of the state of
   polarisation of the (equally prepared) photons. 
  
  Tomography through von Neumann measurements presents an inherent drewback:
   in order to estimate the $d^2-1$ independent parameters of the density matrix, $d+1$ measurements must be realised which means that 
  $d^2+d$ histograms of the counting rate are established, one of them being sacrificed after each of the $d+1$ measurements in order to normalize
   the corresponding probability distribution. From this point of view, the number of counting rates is higher than the number of parameters 
   that characterize the density matrix, which is a form of redundancy, inherent to the tomography through von Neumann measurements.
  
  Besides, it is known that a more general class of measurements exists than
   the von Neumann measurements. This class is represented by the Positive-Operator-Valued
    Measure (POVM) measurements \cite{QCQ}, of which a reduced subset,
     the Projection-Valued Measure (PVM) measurements corresponds to the von Neumann measurements. 
  The most general POVM can be achieved by coupling the system A to an ancilla or assistant B and performing a
   von Neumann measurement on the full system. When both the system and its assistant are 
   qu$d$¥it systems, the full system 
   belongs to a $d^2$ dimensional Hilbert space which makes it possible to measure $d^2$ probabilities during a von Neumann measurement performed on the full system.
    As always, one of the counting rates must be sacrificed in order to normalise the probability distribution so that we are left with $d^2-1$ parameters.
     When the coupling to the assistant and the von Neumann measurement are well-chosen,
      we are able in 
  principle to infer the value of the density matrix of the initial qu$d$it system from the 
  knowledge of those $d^2-1$ parameters, in which case
   the POVM is said to be Informationnally Complete (IC). Obviously, this approach is 
   optimal in the sense that it minimizes the number of counting rates (thus of independent 
   detection processes) 
   that must be realised during the tomographic process.
   
   As it was shown in \cite{WoottersF},
    the PVM approach to tomography can further be optimised regarding redundancy. Optimality according to this particular figure of merit is achieved when
     the $d+1$ bases in which the PVM measurements are performed are ``maximally independent'' or ``minimally overlapping'' so to say when they are mutually unbiased (two
     orthonormal bases of a $d$ dimensional Hilbert space are said to be mutually 
  unbiased bases (MUB's) if whenever we choose one state in the first basis, and a second state in the second basis,
   the modulus squared of their in-product is equal to $1/d$).  It is well-known that,
    when the dimension of the Hilbert space is a prime power, there exists a 
   set of $d+1$ mutually unbiased bases \cite{WoottersF,ivanovic,india}.  This is the case for instance with the bases that
    diagonalize the generalised Pauli operators \cite{india,TD05}. Those unitary operators form a group which is
     a discrete counterpart of the Heisenberg-Weyl group, the group of displacement operators \cite{milburn}, that present numerous
      applications in quantum optics and in signal theory \cite{vourdas}.
     
     A discrete version of the Heisenberg-Weyl group \cite{weyl} also plays an essential 
     role \cite{caves} in the derivation of so-called covariant 
     symmetric-informationally-complete (SIC) POVM's. Such POVM's are intimately associated to a set of $d^2$ minimally overlapping projectors
      onto pure qu$d$it states (the modulus squared of their in-product is now equal to $1/\sqrt{d+1}$). 
      
      We shall compare the respective merits of PVM and POVM tomographic processes 
      in the cases of one and two qubit systems and focus on the
       factorisability of the latter. This question leads us to study the factorisability of 
      discrete Wigner quasidistributions which appear to be a very natural tool in the context
       of one qubit and (factorisable) two qubit tomography. Although in the continuous case,
         the factorisation-property is de facto fulfilled, in discrete dimensions this is not a
          trivial question at all. This why this question raised recently a lot of interest 
          (Refs.¥\cite{Wootters2} to \cite{Vourdas2,Wootters87,Rubin1}).
      
      Generalisations to higher dimensions are discussed in the last section and in the conclusions.

 \section{\label{section1}Tomography of a (single) qubit system.}
     \subsection{\label{section1.1}Optimal PVM approach.}
      The aforementioned traditional approach to tomography for a spin 1/2 particle, 
      that consists of three successive Stern-Gerlach measurements performed along orthogonal directions $X$, $Y$ and $Z$ is optimal among PVM tomographic processes because the 3 corresponding bases are MUB's
       \cite{ivanovic,WoottersF,JSh60}. 
  Considered so, the traditional approach to spin/polarisation tomography is optimal. Actually, the $\sigma$ operators plus the identity constitute
   a discrete counterpart of the displacement operators. Formally,
 they can be defined as follows:
$\sigma_{i,j}=\sqrt{(-)^{i.j}}\sum_{k=0}^1(-)^{k.j}\ket{k+i (mod.2)}\bra{k}$,
 where the labels $i,j,k$ can take values 0 or 1. It is easy to check that, up to a global sign that we presently keep undetermined, $\sigma_{0,0}=Id.,
 \sigma_{1,0}=\sigma_X, \sigma_{1,1}=\sigma_Y$ and $\sigma_{0,1}=\sigma_Z.$
  This set is orthonormal regarding the Trace-norm: ${1 \over d}Tr.\sigma^{\dagger}_{i}\sigma_{j}=\delta_{i,j}$ (here $d=2$) and, like
   the displacement operators in the continuous case, its elements constitute a complete basis of the set of linear operators of the Hilbert space on
    which they act. The Bloch parameters are seen to be in one-to-one correspondence with the qubit Weyl distribution (defined by $w_{i,j}=(1/2)Tr.(\rho.\sigma_{i,j})$),
     which consists, in analogy with its continuous counterpart \cite{weyl}, of 
    the amplitudes of the 
development of the density matrix in terms of the (qubit) displacement operators \cite{vourdas}: $w_{0,0}=1/2$,$w_{1,0}=p_{x}/2$,$w_{1,1}=p_{y}/2$,
  $w_{0,1}=p_{z}/2$. These properties can be generalised to higher dimensions \cite{TD05}, and, when the dimension is a prime power, each PVM measurement in one of the $d+1$ MUB's leads to the estimation of a set of $d-1$ parameters of the 
  Weyl distribution. The measurements performed in different MUB's are independent and the $d+1$ subsets of $d-1$ amplitudes obtained so
   form a partition of (the set of $d^2-1$ independent parameters of) the Weyl distribution and provide a complete tomographic information about the unknown qu$d$it 
  state of the system.
  
   \subsection{\label{section1.2}Optimal POVM approach.}

  It has been shown in the past, on the basis of different theoretical 
  arguments \cite{caves,JRe04,Alla04}, that the optimal IC POVM is symmetric in the sense that it is in one-to-one correspondence with a tetrahedron
   on the Bloch sphere. Intuitively, such tetrahedrons homogenize and minimize the
    informational overlap or redundancy between the four histograms collected during the POVM measurement. 
   Some of such tetrahedrons can be shown to be invariant under the action of the Heisenberg-Weyl group which corresponds
   to so-called Covariant Symmetric Informationnally Complete (SIC) POVM's \cite{caves}. 
   
   Let us now briefly describe how such a POVM measurement could be realised 
   experimentally (we were able recently \cite{durtdu} to implement this POVM measurement on a two qubit NMR quantum computer \cite{ekert}). Let us suppose that we wish to estimate the three parameters
    $\gamma ,\theta ,$and $\varphi $ necessary in order to describe the
unknown state of the qubit $a.$ An ancilla is added to this device as qubit $b$ to form a extending system. 
This device is initially prepared in the state: 
$\rho
_{in}=\rho _{a}\otimes \left\vert 0\right\rangle \left\langle 0\right\vert _{b} $.
 This state differs according to
different input qubits $a.$ In virtue of the \textit{Stinespring-Kraus} theorem\cite{MZ04}, quantum
operations are related to the unitary transformations, a property that we shall now exploit by letting the entire system evolve under unitary
evolution $U$%
\begin{equation}
U=\frac{1}{2}\left( 
\begin{array}{cccc}
e^{i\pi /4}\alpha & \alpha & \beta & -e^{i\pi /4}\beta \\ 
\alpha & -e^{-i\pi /4}\alpha & -e^{-i\pi /4}b & -\beta \\ 
\beta & -e^{i\pi /4}\beta & e^{i\pi /4}\alpha & \alpha \\ 
-e^{-i\pi /4}\beta & -\beta & \alpha & -e^{-i\pi /4}\alpha%
\end{array}%
\right)  \label{U}
\end{equation}%
where $\alpha=\sqrt{1+1/\sqrt{3}},\beta=\sqrt{1-1/\sqrt{3}}$. 

By measuring the full system in a basis that consists of the product of the $a$ and $b$ qubit computational bases, we are able in principle to 
estimate the four parameters
 enlisted on the diagonal of the following matrix:
 
 \label{Th}%
\begin{equation}
\rho _{out}^{Th}=\left( 
\begin{array}{llll}
P_{00} &  &  &  \\ 
& P_{01} &  &  \\ 
&  & P_{10} &  \\ 
&  &  & P_{11}%
\end{array}%
\right)
\end{equation}
such a POVM measurement is informationally complete due to the fact that the $P_{00},P_{01},P_{10},P_{11}$ are in one-to-one correspondence with the Bloch parameters $%
p_{x},p_{y},$ and $p_{z}$ as shows the identity 
\begin{equation}
\left\{ 
\begin{array}{l}
P_{00}=\frac{1}{4}\left[ 1+\frac{1}{\sqrt{3}}(p_{x}+p_{y}+p_{z})\right] \\ 
P_{01}=\frac{1}{4}\left[ 1+\frac{1}{\sqrt{3}}(-p_{x}-p_{y}+p_{z})\right] \\ 
P_{10}=\frac{1}{4}\left[ 1+\frac{1}{\sqrt{3}}(p_{x}-p_{y}-p_{z})\right] \\ 
P_{11}=\frac{1}{4}\left[ 1+\frac{1}{\sqrt{3}}(-p_{x}+p_{y}-p_{z})\right]%
\end{array}%
\right.
\end{equation}

  Actually, $P_{00}$ is the average value of the operator $({1\over 2})(\sigma_{0,0}+
 ({1\over \sqrt 3})(\sigma_{1,0}+\sigma_{0,1}+\sigma_{1,1}))$ which is the projector onto the pure state $\ket{\phi}
   \bra{\phi}$ with $\ket{\phi}=\alpha\ket{0}+\beta^*\ket{1}$ and
    $\alpha=\sqrt{1+{1\over \sqrt 3}} $, 
    $\beta^*=e^{{i \pi\over 4}}\sqrt{1-{1\over \sqrt 3}} $. Under the action of the Pauli group it transforms into a
     projector onto one of the four 
    pure states $\sigma_{i,j}\ket{\phi}$; $i,j:0,1$: $\sigma_{i,j}\ket{\phi}\bra{\phi}
    \sigma_{i,j}=({1\over 2})((1-{1\over \sqrt 3})\sigma_{0,0}+
 ({1\over \sqrt 3})(\sum_{k,l=0}^1(-)^{i.l-j.k}\sigma_{k,l}))$
The signs $(-)^{i.l-j.k}$ reflect the (anti)commutation properties of the Pauli group. So, the four parameters $P_{ij}$ are the average values of
 projectors onto four pure states that are ``Pauli displaced'' of each other.
The in-product between them is equal, in modulus, to $1/\sqrt 3=1/\sqrt{d+1}$, with $d=2$. 
This shows that this POVM is symmetric in the sense that it is in one-to-one correspondence with a tetrahedron
   on the Bloch sphere; as this tetrahedron is invariant under the action of the Heisenberg-Weyl group it is a Covariant Symmetric
    Informationnally Complete (SIC) POVM \cite{caves}. One can show 
    \cite{caves,JRe04,Alla04} that
   such tetrahedrons minimize the informational redundancy between the four 
   collected histograms due to the fact that their angular opening is maximal.
   
   It is worth noting that this POVM possesses another very appealing property which is
    also true in the qutrit case but not in dimensions strictly higher than 3  \cite{gross}:
    the qubit Covariant SIC POVM is 
a direct realisation (up to an additive and a global normalisation constants) of the 
qubit Wigner distribution of the unknown qubit $a$. Indeed, this distribution $W$ ¥is the 
symplectic
 Fourier transform of the Weyl distribution $w$ ¥(already defined by the relation 
 $w_{i,j}=(1/2)Tr.(\rho.\sigma_{i,j})$) which is,
   in the qubit case, equivalent (up to a relabelling of the indices) to its
    double qubit-Hadamard or double qubit-Fourier transform:
  
  \beq{W_{k,l}=(1/2)\sum_{i,j=0}^1(-)^{i.l-j.k}w_{i,j}=((1/\sqrt 2)
  \sum_{i=0}^1(-)^{i.l})((1/\sqrt 2)\sum_{j=0}^1(-)^{-j.k})w_{i,j}.}\label{wign}\eeq
  
  This expression, originally derived by Wootters in a somewhat different form \cite{Wootters87}, is a special case of an expression for a Wigner distribution derived by us
    in prime power dimensions \cite{durtwig}.
  One can check that $P_{k,l}=(1/\sqrt 3)W_{k,l}+(1-1/\sqrt 3)/4.$
   The symplectic Fourier transform is invertible so that once we know the Wigner distribution, we can directly infer the Weyl distribution or,
    equivalently, the Bloch vector of any a priori unknown quantum state. It is worth noting that 
  in order to measure the coefficients $P_{k,l} (W_{k,l})$ we must carry out a
   measurement on the full system (the unknown qubit plus the ancilla), which is the essence
    and novelty of entanglement-assisted quantum tomography \cite{durtdu,qlah}.

        It has been shown that the discrete qubit Wigner distribution directly generalises its
         continuous counterpart \cite{wigner} in the sense that it provides information about the localisation of the qubit
     system in a discrete 2 times 2 phase space \cite{Wootters87}.
      For instance the Wigner distribution of
      the first state of the computational basis (spin up along $Z$) is equal to 
      $W_{k,l}(\ket{0})=(1/2)\delta_{k,0}$, which corresponds to a state located in the ``position'' spin up (along $Z$), and homogeneously spread in ``impulsion'' (in spin along $X$), in accordance with uncertainty relations (see Ref.
      \cite{Rubin1} for an enlightening discussion of discrete uncertainty relations in connection with the Wigner distribution)¥. 
      Similarly, the Wigner distribution of
      the first state of the complementary basis (spin up along $X$) is equal to 
      $W_{k,l}((1/\sqrt 2)(\ket{0}+\ket{1}))=(1/2)\delta_{l,0}$.
      
      We arrived to our formulation of the Wigner distribution \cite{durtwig} by deriving a solution of
       the Mean King's problem \cite{LVaid87,YA01,aravind}, which is not astonishing.
        Indeed, this problem consists of ascertaining the value of the spin 
       of a spinor prepared at random in the 
      $X$, $Y$ and $Z$ bases. The connection between the Wigner distribution and the 
      Mean King problem is the following. Its solution consists of entangling the qubit 
      to another qubit and to measure the full system in a well-chosen quartit basis 
      in such a way that each detector fires with a probability equal to the Wigner distribution of the first qubit. Therefore, knowing 
      to which basis the initial states belong and knowing which of the four ``Wigner'' detectors would fire, 
      we are able to infer what is the value of their spin component \cite{LVaid87,durtwig}. For instance, when the spin is prepared in the 
      $Z$ basis and that the detector corresponding to $W_{1,i}$ ($i=0,1$) fires, we could 
      infer that the initial spin was the state $\ket{1}$ (spin down along $Z$).

\section{\label{section2}Tomography of a two qubit system.}
     \subsection{\label{sect}Optimal PVM approach.}
    The results relative to the qubit case can be generalised to multi-qu$d$it systems whenever 
    $d$ is a prime power  \cite{WoottersF},
     so to say to the case $d=p^m$ with $p$ prime and $m$ a positive 
    integer. In particular, in the present case ($p=m=2$),
     we can define generalised displacement operators according to the definition given in
      \cite{TD05} (see also Ref.\cite{DurtNagler})¥:
    
   \beq
 V^j_i= \sum_{k=0}^{d-1}
  \gamma_{G}^{(( k\oplus_{G} i)\odot_{G} j)}\ket{ k\oplus_{G} i}\bra{  k}
  \label{defV0};i,j:0...d-1, \eeq 
where $\gamma_{G}$ is the $p$th root of unity: $\gamma_{G} =e^{ i .2\pi/p}$, 
the Galois addition $\oplus_{G}$ and the Galois multiplication $\odot_{G}$ ¥are defined by the 
tables given in appendix. The Galois addition 
is by definition equivalent with the addition modulo
  $p$ componentwise. Concretely, this means that if we write the labels (0,1,2,3) in a binary form (0=(0,0),1=(0,1),2=(1,0),3=(1,1)), 
  the Galois addition is defined as follows: if ($i=(i_{a},i_{b})$ and $j=(j_{a},j_{b}))$, 
  then $i\oplus_{G} j=(i_{a}\oplus_{mod 2}j_{a},i_{b}\oplus_{mod 2}j_{b})$.
   The Galois multiplication is distributive relatively to the Galois addition, moreover it is commutative, and there is no divider of 0 (the neutral element for the addition) excepted 0 itself. 
The algebraic structure that is defined by these requirements is a field; in the present case,
 the multiplication table is uniquely defined by these requirements, and by the
  definition of the addition: $0\odot_{G}x=x\odot_{G}0=0;$¥
  $¥1\odot_{G}x=x\odot_{G}1=x;$¥$¥2\odot_{G}3=3\odot_{G}2=1$ and $2\odot_{G}2=3$¥¥.

We can write the 16 4 times 4 displacements operators defined by Eqn.(\ref{defV0})
 as the identity plus 5 sets of 3 operators that are defined as follows.
 The first set consists of the operators $V^j_{i}$ with $i=0$ and $j=1,2,3$, while the other sets
  consist of the operators $V^{(l-1)\odot_{G}i}_{i}$ with $i=0,1,2,3,$ and correspond to the respective choices $l=1,2,3,4$. By direct computation, one can check that we obtain so the five sets 
  $\{ \sigma^a_{z} ,\sigma^b_{z}  ,\sigma^a_{z} .\sigma^b_{z}\}$,
 $\{ \sigma^a_{x} ,\sigma^b_{x}  ,\sigma^a_{x} .\sigma^b_{x}\}$,
 $\{ \sigma^a_{y} ,\sigma^b_{y}  ,\sigma^a_{y} .\sigma^b_{y}\}$,
 $\{ \sigma^a_{x}.\sigma^b_{y} ,\sigma^a_{y}.\sigma^b_{z}  ,\sigma^a_{z} .\sigma^b_{y}\}$ and $\{ \sigma^a_{x}.\sigma^b_{z} ,\sigma^a_{y}.\sigma^b_{x}  ,\sigma^a_{z} .
\sigma^b_{y}\}$ (up to irrelevant global phases). The operators that belong to
 each of these sets commute with each other; moreover, it is possible to multiply each of them by a well-chosen phase in such a way that the 3 operators of 
each set form (together with the identity operator) a commutative group.  The bases
 that simultaneously diagonalize all the operators of such sets are unambiguously defined and are mutually unbiased
\cite{india,TD05} relatively to each other. 
All the properties that we described in the qubit case are still valid in the present case: by performing von Neumann measurements in those 5 MUB's it is possible to estimate 15 parameters 
(5 times (4-1) probabilities) that are in one-to-one correspondence with the Weyl distribution, or equivalently with the coefficients of the density matrix of the system.

The problem is that only the three first bases are factorisable (the common eigenbases of the two
 last operators are actually maximally entangled \cite{india,TD05,sanchez}). This is not astonishing because by taking products of mutually
  unbiased qubit bases we can at most construct 3 MUB's. Of course, it is still possible to obtain
  full tomographic information about the system by measuring each qubit component in the three MUB's (along $X,Y$ and $Z$),
  due to the fact that the displacement operators are factorisable into a product of local displacement
   operators (a very general property, also valid in dimension $p^m$ with $p$ an arbitrary 
   prime and $m$ an
    arbitrary positive integer \cite{TD05}). 
    The problem is that this requires to establish $3^2.2^2=36$ count rates in order to
     estimate 15(+1) parameters, a situation which is far from being optimal. 
     To conclude, it is clearly impossible to conciliate optimality and factorisability in the two qubit case (as well as in the
      two qu$d$it case, when $d=(p^m)^2$) because factorisable MUB's are necessarily products of local MUB's
       ($\sqrt{d^2}=\sqrt{d}.\sqrt{d}$) and, 
      by taking products of mutually
  unbiased bases of the $p^m$ dimensional Hilbert spaces associated to the
 two components of a bipartite system of dimension $d=(p^m)^2$,
 we can at most construct $p^m+1$ (factorisable) MUB's which, for all prime power dimensions, is strictly smaller than $d+1=(p^m)^2+1$.

 \subsection{\label{section2.2}Optimal POVM approach.}
 As in the two qubit case, it is easy to show on the basis of 
 a simple argument that it is impossible to conciliate optimality and factorisability of the POVM or entanglement-assisted tomography in the two qubit case 
 (as well as in the
      two qu$d$it case, when $d=(p^m)^2$). The reason therefore is that when $d=d_{1}.d_{2}$, and that a POVM
       can be splitted into a product of 
      two SIC POVM's which means that we couple the $d_{1}$ ($d_{2}$) dimensional subsytem to a $d_{1}$ 
      ($d_{2}$) dimensional assistant we find $d^2=d_{1}^2.d_{2}^2$ projectors onto 
      factorisable pure states of which the in-products are most often equal in modulus to
       $(1/\sqrt{d_{1}+1}).(1/\sqrt{d_{2}+1})$ but not always; sometimes this
        in-product is equal to $(1/\sqrt{d_{1}+1})$ or $(1/\sqrt{d_{2}+1})$, so that what we
         get is not a SIC POVM, but only an IC POVM which  may not be optimal.
       
        Nevertheless, the number of counting rates necessary in order to realize a tomographic process by a factorisable 
        POVM measurement is optimal and equal to $d^2=d_{1}^2.d_{2}^2$, so that this technique can be considered as
         a good compromise: not fully optimal from the point of view of redundancy but at least factorisable which is very
          appealing regarding the experimental realisability of the tomographic process in the case of separated subsytems
           (for instance in the case of two separated photons entangled in polarisation, a common situation in the laboratory). In the case of two qubits, a factorisable SIC POVM is nearly equivalent (actually, it is equivalent 
           up to a simple renormalisation of the counting rates)  to a direct measurement of the average values of the 
           16 products of the local (qubit) Wigner operators defined in a previous section (Eqn.(\ref{wign}).
            As we are interested in the question of the factorisability of the tomographic process, it is very natural in 
            the present context to ask the following question: Is the product of the two Wigner
             distributions of the subsystems of a bipartite system equivalent to the Wigner
              distribution of the full system? We shall provide certain answers to this question in the following section. 
              
              \section{\label{section3}Factorisability of the discrete Wigner distribution.}
             \subsection{\label{sectionwig}Candidates for the Wigner distribution.}
                
             Before we discuss their factorisability, it is necessary to provide an operational
              definition
               of the discrete Wigner distribution for a finite-state system. We derived such a 
               definition in Ref.\cite{durtwig} where we showed that the recipe for constructing a Wigner distribution associated to a qu$d$it system, 
                with $d=p^m$, was the following:
                
                i) Let us split the set of $d^2$ d times d displacement operators defined by 
                Eqn.(\ref{defV0}) into 
                the identity plus $d+1$ sets of $d-1$ operators that are defined as follows.
 The first set consists of the operators $V^j_{i}$ with $i=0$ and $j=1,...,d-1$, while 
 the other sets
  consist of the operators $V^{(l-1)\odot_{G}i}_{i}$ with $l=1,...,d-1$
   and correspond to the respective choices $l=1,...,d$. In the qubit case,
    each set corresponds to one of the Pauli operators. In the two qubit case, the list
     of these sets was explicitly given in the section \ref{sect}. It can been shown \cite{Durtmutu} that all the operators from a same set commute with each other. 
     
     ii) Let us multiply each operator of a set by a well-chosen phase in such a way that the set of ``renormalised'' operators together with the identity forms a finite group (with $d$ elements).
     
     It is shown in Ref.\cite{TD05} that there are, for each of the $d+1$ sets,
      $d$ possible choices of phases that satisfy this constraint. Moreover these choices are
      equivalent to simple relabellings (in fact translations or Galois-additive shifts)
      of the indices of the states of the MUB in which the operators of the group are simultaneously diagonal.
      
      In the same reference two possible choices are explicitly given, that correspond to the odd and even dimensional cases:
      
      In the odd case, we showed that one among the possible phase choices led to the following definition of the 
           renormalised displacements operators 
           associated to the $i$th group (denoted $ U^i_{l}$, with $i:1,...,d$, $l:0,...,d-1$):
           \beqa \label{elegant}
 U^i_l= (\gamma_{G}^{\ominus_{G}((i-1)\odot_{G} l\odot_{G} l)/_{G}2})V_{l}^{(i-1)\odot_{G} l},\eeqa
  where $/_{G}$ represents the inverse of the field (Galois) multiplication and $2=1\oplus_{G}1$. Actually, this choice
   is particularly attractive and elegant for several reasons and is uniquely defined
                   once we know the operation tables of the field with $p^m$ elements. We shall show that it helps us to answer positively to the problem of the factorisation of the Wigner function for bipartite systems in odd prime power dimensions in a next section.

          In the even case ($d=2^m$) the explicit expressions for the phases are less easy to manipulate, essentially due to the fact that $1\oplus_{G}1=0$ and that we may not divide by 0. 
                Once again, there are $p^m$ ($2^m$ in this case) possible ways to determine the 
   phases $U^j_{1}/V^{(j-1)\odot_{G} l}_{l}$ which are equivalent, up to a relabelling of the states of the 
   corresponding MUB.
   
   A possible choice for the phases was shown to be   
   \beqa U^j_{l}= (\Pi^{m-1}_{n=0, l_{n}\not= 0}i^{(j-1)\odot_{G}2^n\odot_{G}2^n}\gamma_{G}^{(j-1)\odot_{G}2^n\odot_{G}2^{n'}})
 V^{((j-1)\odot_{G} l)}_{l}
 \label{UsurVeven}\eeqa
 where the coefficients $l_{n}$ are unambiguously defined by the $p$-ary (here binary) expansion of $l$, $l=\sum_{k=0}^{m-1}l_{n}2^n$, while 
 $n' $ is the smallest integer strictly larger than $n$ such that $l_{n'}\not= 0$, if it exists, 0 otherwise. 
 
 For $i=0$,  a possible choice of phases corresponds to the relation $U^{0}_{l}=V^{l}_{0}$, in the even and odd cases.
 
 It is worth noting that in both cases the phases are square roots of integer powers of gamma,
 
 $(U^j_{l}/V^{(j-1)\odot_{G} l}_{l})^2=\gamma_{G}^{\ominus_{G}(
  (j-1)\odot_{G} l\odot_{G} l)}$. As a consequence, it is easy to show (see appendix) that $(U^j_{l})^{-1}=(U^j_{l})^{\dagger}=
  U^j_{\ominus_G l}$
  
  iii) In Ref.\cite{durtwig}, the $d^2$ Wigner operators are defined as follows:
  
\beqa \label{wigno}W_{(i_{1},i_{2})}={1\over d}\sum_{m,n=0}^{d-1}
  \gamma_{G}^{\ominus_{G}i_{1}\odot_{G}n \oplus_{G}i_{2}\odot_{G}m}
   (\gamma_{G}^{( m\odot_{G} n)})^{1\over 2} V^{n}_{m}, \eeqa 
  Introducing the more convenient notation 
  $U_{m,n}=(\gamma_{G}^{( m\odot_{G} n)})^{1\over 2} V^{n}_{m}$ (the $U_{m,n}¥$ operators are equivalent
   with the $U^i_{j}¥$ operators previously defined, up to a mere relabelling¥¥),¥ we get
   \beqa W_{(i_{1},i_{2})}={1\over d}(\sum_{m=1,n=0}^{d-1}
  \gamma_{G}^{\ominus_{G}i_{1}\odot_{G}n \oplus_{G}i_{2}\odot_{G}m}
   U_{m,n}). \eeqa
   
    It is worth noting that the Wigner ($W$) operators defined by Eqn.(\ref{wigno}) are Hermitian, due to the identity $U_{m,n}^{\dagger}=U_{m,n}^{-1}=U_{\ominus_{G}m,\ominus_{G}n}$. In appendix, we prove that they are ``acceptable'' candidates for discrete
    Wigner distributions according to the criteria introduced by 
                Wootters in his seminal paper of '87 \cite{Wootters87}, which means that (a) their Trace is equal to 1, (b) they
                 are orthonormalised (to $d$) operators (under the Trace norm) so that they form a basis of the set of linear operators, 
                (c) that if we consider any set of parallel lines in the phase space, the average of the Wigner operators along one of 
                those lines is equal to a projector onto a pure state, and the averages taken along different parallel lines are projectors onto 
                mutually orthogonal states. In appendix we prove that the last relation is valid in odd and even prime power dimensions as well and we 
                also show that the sets of $d$ orthogonal states that are associated to different
                 directions (there are $d+1$ non parallel directions in the $d$ times $d$ phase 
                 space) form MUB's (in accordance with the prediction made in Ref.\cite{discretewigner})¥.
                 
                 The Wigner distribution is then nothing else than the set of the $d^2$ amplitudes that we obtain when we develop
                  the density matrix of a qu$d$it (with $d=p^m$) in the basis provided by the Wigner operators:
                 
                 $w_{(i_{1},i_{2})}=Tr.(\rho.W_{(i_{1},i_{2})})$ 
                 
                 As a consequence of the property $c$ the marginals of this distribution along any axis of the phase space are equal to transition probabilities to states of the corresponding MUB.
                 
                 \subsection{\label{section3.2}Factorisability of the two and three qubit Wigner distribution.}

                In Ref.\cite{durtwig}, we showed that 
                $W_{(i_{1},i_{2})}=(V_{i_{1}}^{i_{2}})^{-1}W_{(0,0)}V_{i_{1}}^{i_{2}}$
                , and we also showed in Ref.\cite{TD05} that the displacement operators always factorise into products of local displacement operators. Therefore, in order to prove the factorisability of the Wigner operators, it is sufficient to prove that 
                $W_{(0,0)}$ factorises.
                As we mentioned in the previous section, the Wigner operators are not uniquely defined according to our definition.
                 For instance, in the qubit case we are free to change the sign of the operators $\sigma_{x}$, $\sigma_{y}$, and $\sigma_{z}$ according to our convenience
              in the definition (\ref{wign}). This change of sign is equivalent to a relabelling of the states of the 
             associated MUB's which has no important physical consequence. Therefore $2^3$ acceptable Wigner
              distributions exist in the qubit case. In the two qubit case, we are free to choose arbitrarily the sign of
               two operators in each of the 5 families of 3 operators that we 
             defined in the section \ref{sect}, which corresponds to $4^5$ acceptable Wigner distributions.
              There are thus $(2^3)^2$ possible products of two qubit Wigner distributions and 
             $4^5$ possible two-qubit (quartit distributions). Once we have 
             chosen the signs of the three sigma operators ($x,y,z$) 
             of the first qubit, the corresponding signs at the level of the second qubit must be the same
              an even number
              of times (0 or 2 times) in order that the product of the Wigner distributions is an acceptable 
              two qubit distribution. For instance, if we choose the phase $+$ for the $a$ sigma operators ($x,y,z$) and
               the phase $-$ for the $b$ sigma operators ($x,y,z$), we obtain the products  $\{ +\sigma^a_{z} ,
               -\sigma^b_{z}  ,-\sigma^a_{z} .\sigma^b_{z}\}$,
 $\{ +\sigma^a_{x} ,-\sigma^b_{x}  ,-\sigma^a_{x} .\sigma^b_{x}\}$,
 $\{ +\sigma^a_{y} ,-\sigma^b_{y}  ,-\sigma^a_{y} .\sigma^b_{y}\}$,
 $\{ -\sigma^a_{x}.\sigma^b_{y} ,-\sigma^a_{y}.\sigma^b_{z}  ,-\sigma^a_{z} .\sigma^b_{y}\}$ and 
 $\{ -\sigma^a_{x}.\sigma^b_{z} ,-\sigma^a_{y}.\sigma^b_{x}  ,-\sigma^a_{z}\}$.
 Each of these families (¥¥together with the identity) forms a commuting group, so that the corresponding operator $W_{(0,0)}$,
  which is the sum of those 16 operators factorises:
 \beqa W_{(0,0)}={1\over 4}(+Id.+\sigma^a_{z} 
               -\sigma^b_{z}  -\sigma^a_{z} .\sigma^b_{z}+\sigma^a_{x} -\sigma^b_{x}  -\sigma^a_{x} .\sigma^b_{x}+
               \nonumber 
               \\ \sigma^a_{y} -\sigma^b_{y}  -\sigma^a_{y} .\sigma^b_{y}-\sigma^a_{x}.\sigma^b_{y} -\sigma^a_{y}.\sigma^b_{z}  -\sigma^a_{z}.\sigma^b_{y}-\sigma^a_{x}.\sigma^b_{z} -\sigma^a_{y}.\sigma^b_{x}  -\sigma^a_{z}.\sigma^b_{y} ) \nonumber 
               \\= {1\over 2}(+Id.^a+\sigma^a_{x} 
               +\sigma^a_{y}  +\sigma^a_{z}).{1\over 2}(+Id.^b-\sigma^b_{x} 
               -\sigma^b_{y}  -\sigma^b_{z}).\eeqa
  As a consequence, all Wigner operators factorize too. There are obviously $2^3.4=2^5$ similar ways to derive factorisable two-qubit Wigner 
  distributions, so that fifty percent of the products of qubit Wigner distributions provide an acceptable quartit Wigner distribution (among the $4^5$ acceptable Wigner distributions).
 
It is now easy to show that 
                    a three qubit Wigner distribution never factorises into the product of three qubit Wigner distributions.
                     Essentially this is due to the fact that it is impossible to find three ordered triplets of plus or minus signs 
                     that would be the same an even number of times (0 or 2 times) TWO BY TWO. 
                     
                     It is important to note that these results are still valid in the approach followed in 
                     Ref.\cite{discretewigner}. This (more axiomatic and geometrical)
                       ¥approach is more general because it allows more flexibility in the way to attribute one element of the Galois field to one label of the basis states
                        of the computational and dual bases. Implicitly in our approach, the $m$uple of integers comprised between $0$ and $p-1$ assigned to 
                        each computational state expresses the development of the corresponding element of the Galois field in a field ¥basis (field bases are defined in Ref.\cite{discretewigner})¥
                       that contains 1 (the neutral element for the multiplication) as first element. Similarly, our choice of the dual basis is such that the MUB associated to diagonal straightlines factorises into local MUB's associated to local 
                       diagonal straight lines \cite{TD05}. Anyhow,  the MUB's are exactly the same in both approaches, independently on the choices of field bases,
                        because the splitting of the Heisenberg-Weyl into commuting sub-groups is unambiguously defined up to relabellings
                     once
                     we associate $\sigma_{X}¥¥$ operators to horizontal translations (shifts of the labels of the first MUB) 
                     ¥and ¥¥$\sigma_{Z}¥¥$ operators
                      to vertical translations (shifts of the labels of the second MUB). In our approach, different choices of phase conventions for the subgroup
                       of the Heisenberg-Weyl group lead to different labellings of the corresponding MUB while in the approach of  Ref.\cite{discretewigner} this labelling is arbitrarily imposed. Anyhow, 
                       there are in both cases 
                    $(¥¥d)¥^{d+1¥}¥$ possible phase conventions in dimension $d$ once the field bases are chosen. There are thus in both approaches 
                     $(¥¥d^{2¥})¥^{d^2+1¥}¥$ possible Wigner operators in dimension $d^2$ and ¥$d^{2(¥¥d+1)¥}¥$ possible 
                     factorisable Wigner operators, and those operators are the same because they can always be written
                      (¥up to additive and normalisation constants) as sums of projectors onto $d+1$ states from different MUB's \cite{discretewigner}, and the MUB's are the same in both approaches. 
                     
                      One can check for instance that one of the two factorisable Wigner distributions derived in Ref.\cite{discretewigner}
                       in the case $d=4$ 
                       corresponds to choosing the phases $(+,+,+)¥$ for the ¥$a$ sigma operators ($x,y,z$) and
               the phases $(+,-,+)¥$ for the $b$ sigma operators ($x,y,z$). The second one is obtained similarly but with a 
               permutation of the roles of $a$ and $b$. The 32 operators that we derived can be obtained from those two operators 
               by performing at most one local rotation of 180 degrees around the $X$, $Y$ or $Z$ axis.
                     
                     \subsection{\label{odd}Factorisability of the two qu$d$it Wigner distribution with $d=p^m$ and $p$ odd.}
               
             We shall now prove the following result: provided we apply in odd prime power dimension the particular phase convention
             (\ref{elegant}) and define the field with $d^2=(p^m)^2$ elements as the quadratic extension of the field 
             with $d=p^m$ elements (this technique was succesfully applied by us in the past in order to solve the 
             Mean King's problem in prime power dimensions \cite{durtwig} ), the Wigner distribution of a bipartite
                    $p^{2m}$-dimensional system naturally factorises into a product of local Wigner distributions for the two 
                    $p^{m}$-dimensional subsystems. 
                    
                    Before we do so, we must define the quadratic extension of a field. Let us denote $i_{a}$ ($i_{b}$) the elements of
                     the field with $p^m$ elements associated to the $a$ ($b$) subsystems. Their quadratic extension is a field 
                    associated to the bipartite system $a-b$ the elements of which we denote ($i_{a}$ ,$i_{b}$). 
                    As always, its addition (denoted $\oplus\oplus_{G}$) is the addition componentwise:  
                    
                    ($i_{a}$ ,$i_{b}) \oplus\oplus_{G}(j_{a},j_{b}$)=($i_{a}\oplus_{G}j_{a}$ ,$i_{b}\oplus_{G}j_{b}) .$
                    
                    In particular, $2=(1_{a}$ ,$0_{b}) \oplus\oplus_{G}(1_{a},0_{b})=(1_{a} \oplus_{G}1_{a},0_{b})$=$¥(2_{a},0_{b})$¥¥¥
             
             All what we need to know about the extended multiplication rule 
             (denoted $\odot\odot_{G}$) is that it is commutative and distributive relatively to the extended addition,
              that
               
               ($i_{a} ,0) \odot\odot_{G}(j_{a},0$)=($i_{a}\odot_{G}j_{a}$ ,0) ,
               
               ($i_{a} ,0) \odot\odot_{G}(0,j_{b}$)=($0,i_{a}\odot_{G}j_{b}$ ), and finally that
               
              $ (0,i_{b})\odot\odot_{G}(0,j_{b}$)=($i_{b}\odot_{G}j_{b}\odot_{G} R$, 
              $i_{b}\odot_{G}j_{b}\odot_{G} Q$), with $R$ and $Q$ elements of the
               field with $p^m$ elements, and $R$ different from 0 (otherwise this extension
                would not form a field). Those properties are very similar to those met in the case of the complex field which is the quadratic extension of the real (infinite) field. 
               
               It is instructive to note that as  $2=(2_{a},0_{b})$, 
               its inverse (relatively to the extended multiplication) is equal to 
               $¥(2^{-1}_{a},0_{b})$¥¥¥¥, where $2^{-1}$ represents the inverse of 2 in 
               the non-extended field with $p^m$ elements.¥¥

               The Wigner operators of the composite system can now be written according to 
               Eqn.(\ref{wigno}), and their factorisability is easily established:
               
               \beqa &&W_{((i^1_{a} ,i^1_{b}),(i^2_{a} ,i^2_{b}))}={1\over d^2}\sum_{(m_{a} ,m_{b},n_{a} ,n_{b})=0}^{d-1}
  \gamma_{G}^{\ominus\ominus_{G}(i^1_{a} ,i^1_{b})\odot\odot_{G}(n_{a} ,n_{b})\oplus\oplus_{G}(i^2_{a} ,i^2_{b})
  \odot\odot_{G}(m_{a} ,m_{b})}
   (\gamma_{G}^{((m_{a} ,m_{b})\odot\odot_{G} (n_{a} ,n_{b})})^{1\over 2} V^{(n_{a} ,n_{b})}_{(m_{a} ,m_{b})}\qquad\nonumber\\
  && ={1\over d^2}\sum_{(m_{a} ,m_{b}),(n_{a} ,n_{b})=0}^{d-1}
  \gamma_{G}^{\ominus\ominus_{G}(i^1_{a} ,i^1_{b})\odot\odot_{G}(n_{a} ,n_{b})\oplus\oplus_{G}(i^2_{a} ,i^2_{b})
  \odot\odot_{G}(m_{a} ,m_{b})}
  \gamma_{G}^{ ((m_{a} ,m_{b})\odot\odot_{G} (n_{a} ,n_{b})//_{G}2)}¥). \qquad\nonumber\\
   && \sum_{k_{a} ,k_{b}=0}^{d-1}
  \gamma_{G}^{(( (k_{a} ,k_{b})\odot\odot_{G}  (n_{a} ,n_{b}))}
  \ket{  (k_{a} ,k_{b})\oplus\oplus_{G} (m_{a} ,m_{b})}\bra{   (k_{a} ,k_{b})}
  \nonumber\\
   &&={1\over d^2}\sum_{(m_{a} ,m_{b},n_{a} ,n_{b})=0}^{d-1}
  \gamma_{G}^{\ominus_{G}(i^1_{a}\odot_{G} n_{a}\oplus_{G}i^1_{b}\odot_{G}R\odot_{G}n_{b})\oplus_{G}
  (i^2_{a}\odot_{G} m_{a}\oplus_{G}i^2_{b}\odot_{G}R\odot_{G}m_{b})}
  \gamma_{G}^{ (m_{a}\odot_{G} n_{a}\oplus_{G} (m_{b}\odot_{G}
  R\odot_{G} n_{b})\odot_{G} 2^{-1}}. \qquad\nonumber\\
    &&\sum_{k_{a} ,k_{b}=0}^{d-1}
  \gamma_{G}^{ ((k_{a} \odot_{G}  n_{a})\oplus_{G} (k_{b} \odot_{G}R\odot_{G}  n_{b}))}
  \ket{  k_{a} \oplus_{G} m_{a}}\bra{   k_{a} }\ket{  k_{b}\oplus_{G} m_{b}}\bra{   k_{b}}\qquad\nonumber\\
  &&=W^a_{(i^1_{a} ,i^2_{a})}.W^b_{(i^1_{b} ,R\odot_{G}i^2_{b})}
  \eeqa

             \section{Conclusions.}
             
              At first sight, the tomography of single and two qubit systems seems to be a trivial question.
               From the previous treatment we see that if we analyze the problem at the light of two criteria
                (optimality in the sense of minimal redundancy and factorisability) the problem is surprisingly rich.
                 In particular it motivates the interest of studying the possibility to 
                 factorise the Wigner distribution 
                 of a discrete composite system, a question that recently attracted 
                 an increasing interest (Refs.¥\cite{Wootters2} to \cite{Vourdas2}).
                 
                 Besides the question of tomography of composite 
                 systems, 
                 the main reason
                  therefore is that the phase space structure \cite{Wootters2,Planat} of composite systems 
                  is not necessarily factorisable (for instance the two
                   qubit straight line of slope 2 of the 16 dimensional phase space contains the 
                 4 couples $(0,0)_{a,b},(1,2)_{a,b},(2,3)_{a,b}¥$ and (3,1); it is obviously not the Cartesian
                  product of two qubit straight lines because 
                  ¥$(0_{a}¥,0_{a}¥)=(0,0)_{a}¥,(1_{a},2_{a})=(0,1)_{a},(2_{a},3_{a})=(1,1)_{a}¥$ and
                   $(3_{a},1_{a})¥ =(1,0)_{a}$¥ ). Actually, the existence of non-factorisable lines is directly related to the existence of entangled, non-factorisable MUB's and is an unavoidable feature of composite systems of prime power dimensions, for dimensional reasons similar to those that we explained at the end of section \ref{sect}. 
                   This motivates the quest for global (non necessarily factorisable)
                   phase space approaches, an example of which is provided by our 
                   displacement operators and our Wigner distribution. 
                   Although the phase space of the composite system is not factorisable (it is not the Cartesian product of the phase spaces of the composite systems), it could be that Weyl or Wigner operators nevertheless factorize, which is the object of the present paper. 
                   ¥Of course, the product of local Wigner functions can always provide a full
                    tomographic representation of the state of a composite system,
                     but this picture is naturally linked to a Cartesian splitting of the full phase 
                     space, which is not as rich as a global description of the full phase space. 
                     The affine structure of the full space phase (which is intimately related to the 
                     underlying field with $p^m$ elements¥) is lost whenever we replace it by 
                     the Cartesian product of the local phase spaces (as is done in Ref.\cite{Wootters87}) ¥as showed our example at the beginning of this section.
                   ¥
               
               There exist several approaches to the problem (Refs.¥\cite{Wootters2} to 
               \cite{Vourdas2,Wootters87,Rubin2,Rubin1}),
                and most often those approaches have a pronounced geometrical 
               flavour in the sense that they aim at deriving potential candidates for the Wigner distribution 
             from general considerations about the structure of the 
             $d$ times $d$ phase space (an excellent survey of the phase space approach
              is given in the introduction of Ref.\cite{intro})¥. Our approach is slightly different from the beginning because
              we postulate from the beginning (and this is an educated guess) ¥the ``algebraic'' expression of
                  the Weyl and Wigner operators 
                 (or phase-point operators following the terminology introduced by Wootters in \cite{Wootters87}). 
                 It seems nevertheless that our approach captures the essential features of the more general, geometric,
                  approach. For instance, it is also true in our approach that straight lines of the phase space correspond to states, and that the
                   states associated to parallel lines form orthogonal bases, while different orientations correspond to MUB's
                    \cite{Wootters87,discretewigner}. 
                     Our results about the factorisability of Wigner distributions are still partial results, and they
                      directly raise a question the answer of which is out of the scope of the present paper:

                     Is it possible to factorise the Wigner distribution of a $(2^m)^2$ dimensional system into a product of two (local)
                      Wigner distributions of $(2^m)$ dimensional subsystems when $m$ is strictly larger than 1?
                     
                     Finally, it would also be interesting to investigate the factorisability of Wigner operators 
                     in odd dimensions (in which case
                      it has been shown
                     that many results valid in odd prime power dimensions can be transfered nearly integrally  
                     \cite{gross,Appleby,davidsThesis,cohendet}). For instance the definitions of the Weyl 
                     (¥¥\ref{defV0}) and Wigner operators (\ref{wigno}) are still operational when we replace the Galois operations by the modulo $d$ operations and the $p$th root of unity 
                     ¥¥$\gamma_{G}¥$ by the $d$th root of unity. Factorisation is still possible
                      in this case. For instance when $d=15=3.5$, we can write $m_{a,b}=5.m_{a}+3.m_{b} (modulo 15)¥¥¥¥$, where 
                     $0\leq m_{a,b}\leq 14 $, $0\leq m_{a}\leq 2 $ and $0\leq m_{b}\leq 4 $. Then 
                     factorisation is ensured by the identities $m_{a,b}+n_{a,b} (modulo 15)=5.(¥¥m_{a}+n_{a} (modulo 3))+3.(¥¥m_{b}+n_{b} (modulo 5))¥ $ and $exp^{{i.2\pi¥m_{a,b}.n_{a,b}\over 15}}¥$=$exp^{{i.2\pi¥m_{a}.n_{a}\over 3}}¥$.$exp^{{i.2\pi¥m_{b}.n_{b}\over 5}}¥$.¥¥¥¥¥¥

                     Prime power dimensions remain exceptional anyhow because Galois fields
                      with $d$ elements are known
                       to exist only when $d$ is a prime power. Our guess is that a
                        set of $d+1$ MUB's only 
                       exists in prime power dimensions \cite{Archer}, due to the fact
                        that they seem to be 
                      closely related to the existence of finite fields and finite projective spaces 
                      \cite{Wootters2,Planat,grassl} (finite affine spaces with $d^2$ elements do not 
                      exist when $d=6$ ¥\cite{tarry}  or $d=10$ \cite{lam} and it is
                       conjectured that they exist only when $d$ is
                       a prime power ), but this question is still open.
                       
                       Last but not least, it is worth to mention the approach to the problem that was developed by Rubin and Pittenger 
                        \cite{Rubin2,Rubin1}. This approach is also algebraical but 
                        the authors make use of sophisticated techniques like
                         field extensions of arbitrary order $m$ (in the treatment of odd 
                         dimensions we limited ourselves to quadratic extensions),
                          which enabled them to prove
                         that it is possible to find, when the dimension is an odd prime power
                          ($d$= $p^m$ with  $p$ prime and ODD),
                           Wigner distributions that would be 
                     the product of $m$ local Wigner distributions.
                      Our result of the section \ref{odd} allow
                      us to answer positively whenever $m=2^n, n=1,2,3...$.
                      
                       The authors also established in Ref. \cite{Rubin1} the factorisability of the two qubit Wigner distribution but their
                        method did not allow them to treat systems composed of more than two qubits. As we showed in this paper in the even case ($p=2$) we know that factorisability is not possible when $m=3$ which establishes a clear distinction between the even and odd (prime power) ¥dimensions.
                      
                      Another adavantage of their approach is that they are able to estimate the degree of separability of MUB's 
                      states \cite{Rubin2} in arbitrary prime power dimensions (this reference was kindly drawn to my attention by the authors).
                       It seems that our approaches are closely related 
                       (for instance, in order to establish the separability
                        of Wigner distributions \cite{Rubin1}, the authors also made use of the relative freedom in the assignment of phases to commuting operators of a same subgroup).
                         In the introduction of Ref.\cite{Rubin2}, Rubin and Pittenger wrote, 
                         relative to the approach of
                         Wootters {\it et al.} that {\it ``Although the motivations of the two approaches appear to be quite different, they require the same mathematical tools and appear to lead to the same results. An 
                        interesting question is the interrelationship between the two approaches.''} This is certainly true concerning our approach and theirs.

   \medskip                      
   \leftline{\large \bf Acknowledgment}
\medskip T.D. acknowledges
a Postdoctoral Fellowship of the Fonds voor Wetenschappelijke Onderzoek,
Vlaanderen and also support from the
IUAP programme of the Belgian government, the grant V-18, and the Solvay Institutes for
 Physics and Chemistry. Support from the Quantum Lah at N.U.S. (among others comments of B-G Englert) is acknowledged.

\leftline{\large \bf Appendix 1: ``Acceptability'' of the Wigner operators.}
   Let us consider the Wigner operators defined by Eqn.(\ref{wigno}). We shall now show that (a) their Trace is equal to 1, (b) they
                 are orthogonal with each other and normalised to $d$ (under the Trace norm) (c) that if we consider any set of parallel lines in the phase space, the average of the Wigner operators along one of 
                those lines is equal to a projector onto a pure state, and the averages taken along different parallel lines are projectors onto 
                mutually orthogonal states.
                
                Three identities, that were derived in Ref.\cite{TD05}, will be helpful in the establishment of the proofs:
                \beq\label{identi1}\sum_{j=0}^{d-1} \gamma_{G}^{ (j\odot_{G} i)}=d\delta_{i,0}\eeq 

                \beq\label{identi2}\gamma_G^{i}\cdot\gamma_G^{j}=\gamma _{G} ^{(i\oplus_G j)}\eeq
                
                \beqa
 V^j_i.V^k_l=\gamma^{\ominus_{G}(i\odot_{G} k)} V^{j\oplus_G k}_{i\oplus_{G} l}.
 \label{compo} \eeqa
                
                With the help of the identity (\ref{identi1}) and on the basis of the definition  (\ref{defV0}), we get that  
                $tr.V^j_i=\sum_{k,k'=0}^{d-1}
  \gamma_{G}^{(( k\oplus_{G} i)\odot_{G} j)}\delta_{k\oplus_{G} i,k'}\delta_{k,k'}$$=d.\delta_{i,0}.\delta_{j,0}$. It is easy to show that the $V$ operators defined in (\ref{defV0})
       are unitary with $(V^j_i)^+=(V^j_i)^{-1}=\gamma^{\ominus_{G}(i\odot_{G} j)}
       V^{\ominus_G i}_{\ominus_G j}$.
   Making use of the composition law (\ref{compo}), it is easy to show that the $V$ operators are orthogonal relatively to the trace norm: 
   and that $tr.V^j_i=N.\delta_{i,0}.\delta_{j,0}$.

               Hence, we can derive the property (a):
               
                $Tr.(W_{(i_{1},i_{2})})=Tr.({1\over d}\sum_{m,n=0}^{d-1}
  \gamma_{G}^{\ominus_{G}i_{1}\odot_{G}n \oplus_{G}i_{2}\odot_{G}m}
   (\gamma_{G}^{( m\odot_{G} n)})^{1\over 2} V^{n}_{m})$
   
   =${1\over d}\sum_{m,n=0}^{d-1}
  \gamma_{G}^{\ominus_{G}i_{1}\odot_{G}n \oplus_{G}i_{2}\odot_{G}m}
   (\gamma_{G}^{( m\odot_{G} n)})^{1\over 2} Tr.(V^{n}_{m})$
   
   =${1\over d}\sum_{m,n=0}^{d-1}
  \gamma_{G}^{\ominus_{G}i_{1}\odot_{G}n \oplus_{G}i_{2}\odot_{G}m}
   (\gamma_{G}^{( m\odot_{G} n)})^{1\over 2}.d. \delta_{m,0}.\delta_{n,0})$=${1\over d}\sum_{m,n=0}^{d-1}
  .d=1.$
  
  In order to prove the property (b), we should firstly note that $(U^j_{l})^{-1}=(U^j_{l})^{\dagger}=
  U^j_{\ominus_G l}$ a direct consequence of the Eqn.(\ref{compo}) and of the identity 
  $(U^j_{l}/V^{(j-1)\odot_{G} l}_{l})^2=\gamma_{G}^{\ominus_{G}(
  (j-1)\odot_{G} l\odot_{G} l)}$.
Besides, the $U$ operators, like the $V$ operators are orthogonal relatively to the trace norm, so that  
$Tr.(U^{\dagger}_{m,n}.U_{m',n'})=d.\delta_{m,m'}.\delta_{n,n'}$.
   
  Henceforth, $ Tr.(W^{\dagger}_{(i_{1},i_{2})}.W_{(i'_{1},i'_{2})})={1\over d^2}(\sum_{m=1,n=0}^{d-1}
  \gamma_{G}^{\oplus_{G}i_{1}\odot_{G}n \ominus_{G}i_{2}\odot_{G}m}
  \gamma_{G}^{\ominus_{G}i'_{1}\odot_{G}n \oplus_{G}i'_{2}\odot_{G}m})$
  
  =$
   \delta_{(i_{1},i'_{1})}\delta_{(i_{2},i'_{2})}$, where we applied twice the identity (\ref{identi1}).

  In order to prove the property (c), it is useful to recall the transformation law of the $U$ operators that was 
  established in Ref.\cite{durtwig}:
  
  $U_{m,n}(0)=
  {(\gamma_{G}^{ (\ominus_{G} ((i-1)\odot_{G}  m\ominus_{G}n)\odot_{G} m )})^{{1\over 2}}\over 
  (\gamma_{G}^{ \ominus_{G} ((i-1)\odot_{G}  m\odot_{G} m )})^{{1\over 2}}(\gamma_{G}^{ \oplus_{G} 
  (  m\odot_{G} n )})^{{1\over 2}}}.U_{\ominus_{G}n\oplus_{G}(i-1)\odot_{G}m,m}(i)$, where

   $U_{m,n}(0)=(\gamma_{G}^{ \oplus_{G} (  m\odot_{G} n )})^{{1\over 2}}\sum_{k=0}^{d-1}
    \gamma_{G}^{(( k\oplus_{G} m)\odot_{G} n)} \ket{ e_{k\oplus_{G} m}^0}\bra{  e_{k}^0}$ and
    
     $U_{m,n}(i)=(\gamma_{G}^{ \oplus_{G} 
  (  m\odot_{G} n )})^{{1\over 2}}
  \sum_{k=0}^{d-1} \gamma_{G}^{(( k\oplus_{G} m)\odot_{G} n)}\ket{ e_{k\oplus_{G} m}^i}
  \bra{  e_{k}^i}; i:1...N$. Here, the symbol $\ket{  e_{k}^i}$ represents the $k$th state of the $i$th MUB ($i:0,...d$).
  It is worth noting that in odd prime power dimensions the phase factor ${(\gamma_{G}^{ (\ominus_{G} ((i-1)\odot_{G}  m\ominus_{G}n)\odot_{G} m )})^{{1\over 2}}\over 
  (\gamma_{G}^{ \ominus_{G} ((i-1)\odot_{G}  m\odot_{G} m )})^{{1\over 2}}(\gamma_{G}^{ \oplus_{G} 
  (  m\odot_{G} n )})^{{1\over 2}}}$ is always equal to 1 so that the Wigner operators are invariant in all MUB's (up to a relabelling) \cite{durtwig}.
   The present treatment is also valid in even prime power dimensions.
  
   Let us now evaluate the value of the averaged sum of the Wigner operators (expressed in the computational basis) along a vertical straight line of the phase space:

   $d^{-1}\sum_{i_{2}=0}^{d-1}W^{0}_{(i_{1},i_{2})}=
   d^{-2}\sum_{i_{2}=0}^{d-1} \sum_{m,n=0}^{d-1}$
    $(\gamma_{G}^{\ominus_{G}(i_{1}\odot_{G} n\oplus_{G}m\odot_{G} i_{2})})$
       $    U_{m,n}(0)$
       
        $=
    d^{-2}.\sum_{m,n=0}^{d-1}$
    $d.\delta_{m,0}(\gamma_{G}^{\ominus_{G}
    (i_{1}\odot_{G} n)})$
     $    U_{m,n}(0)$ $=
    d^{-1}.\sum_{m,n=0}^{d-1}$
    $(\gamma_{G}^{\ominus_{G}
    (i_{1}\odot_{G} n)})$ $    U_{0,n}(0)$
   $=
    \sum_{m,n,k=0}^{d-1}$
    $d^{-1}.(\gamma_{G}^{\ominus_{G}
    (i_{1}\odot_{G} n)})$$(\gamma_{G}^{\oplus_{G}
    (k\odot_{G} n)})$ $    \ket{  e^{0}_{k}}\bra{  e^{0}_{k}}$ 
    $=
    d^{-1}.\sum_{m,n,k=0}^{d-1}$
    $d.\delta_{i_{1},k}$ $    \ket{  e^{0}_{k}}\bra{  e^{0}_{k}}$

    $=
    \ket{  e^{0}_{i_{1}}}\bra{  e^{0}_{i_{1}}} .$

  Let us finally evaluate the value of the averaged sum of the Wigner operators along a non-vertical straight line of the phase space
   (of slope $k-1, k:1...d$); this sum is equal to $ d^{-2}.\sum_{\alpha,m,n=0...d-1 }
  \gamma_{G}^{\ominus_{G}i_{1}\odot_{G}n \oplus_{G}i_{2}\odot_{G}m}U_{m,n}(0)$ where $, i_{1}=\alpha_{0}+\alpha,i_{2}=(k-1)\odot_{G}\alpha$;
  
  we can rewrite it in the form

  $d^{-2}.\sum_{\beta,m,n=0...d-1 }
  \gamma_{G}^{\ominus_{G}i'_{1}\odot_{G}n' \oplus_{G}\beta\odot_{G}m'}.
         {(\gamma_{G}^{ (\ominus_{G} ((k-1)\odot_{G}  m\ominus_{G}n)\odot_{G} m )})^{{1\over 2}}\over 
  (\gamma_{G}^{ \ominus_{G} ((k-1)\odot_{G}  m\odot_{G} m )})^{{1\over 2}}(\gamma_{G}^{ \oplus_{G} 
  (  m\odot_{G} n )})^{{1\over 2}}}.U_{m',n'}(k)
    $ where $i'_{1}=(k-1)\odot_{G}i_{1}\ominus_{G}i_{2}=(k-1)\odot_{G}\alpha_{0},    i'_{2}=i_{1}=\beta, m'=(k-1)\odot_{G}  m\ominus_{G}n$ and $n'=m$.
    summing firstly over $\beta$ and making use of the identity (\ref{identi1}), we obtain
     a factor $d.\delta_{m',0}$; now,
     when $m'=0$ then ${(\gamma_{G}^{ (\ominus_{G} ((k-1)\odot_{G}  m\ominus_{G}n)\odot_{G} m )})^{{1\over 2}}\over 
  (\gamma_{G}^{ \ominus_{G} ((k-1)\odot_{G}  m\odot_{G} m )})^{{1\over 2}}(\gamma_{G}^{ \oplus_{G} 
  (  m\odot_{G} n )})^{{1\over 2}}}=1$; the sum reduces thus to 
  
  $d^{-1}.\sum_{n'=0...d-1 }\gamma_{G}^{\ominus_{G}i'_{1}\odot_{G}n' }.U_{0,n'}(k);$
   this is the sum of the Wigner operators along a vertical line, with the operators rewritten in
    the $k$th MUB; in virtue of the previous result, it is equal to 
    $\ket{  e^{k}_{i_{1}}}\bra{  e^{k}_{i_{1}}} $, the projector onto the $i_{1}$th state of the $k$th MUB.
   
   \newpage

  \leftline{\large \bf Appendix 2: Galois operations in dimension 4}
  
  \medskip

 \begin{table}
 \begin{tabular}{c||c|c|c|c}
\hline $\odot_{G}$  & $0$ & $1$ & $2$  & $3$\\
\hline \hline  0 & 0 & $0$ & $0$  & $0$ \\
1 & $0$ & $1$ & $2$  & $3$\\
2 & $0$ & $2$ & $3$  & $1$\\
3 & $0$ & $3$ & $1$  & $2$ \\
 \hline
\end{tabular}
\caption{The field (Galois) multiplication in dimension 4. }\label{tab1}
 
\begin{tabular}{c||c|c|c|c}
\hline $\oplus_{G}$  & $0$ & $1$ & $2$  & $3$\\
\hline \hline  0 & 0 & $1$ & $2$  & $3$ \\
1 & $1$ & $0$ & $3$  & $2$\\
2 & $2$ & $3$ & $0$  & $1$\\
3 & $3$ & $2$ & $1$  & $0$ \\
 \hline
\end{tabular}
\caption{The field (Galois) addition in dimension 4. }\label{tab2}
\end{table}

\end{document}